# Anomalous quantized plateaus in two-dimensional electron gas with gate confinement

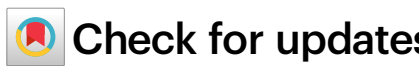

Jiaojie Yan[1], Yijia Wu[1], Shuai Yuan[1], Xiao Liu[1], L. N. Pfeiffer[2], K. W. West[2], Yang Liu[1], Hailong Fu[3], X. C. Xie[1,4] & Xi Lin[1,4,5]

Quantum information can be coded by the topologically protected edges of fractional quantum Hall (FQH) states. Investigation on FQH edges in the hope of searching and utilizing non-Abelian statistics has been a focused challenge for years. Manipulating the edges, e.g. to bring edges close to each other or to separate edges spatially, is a common and essential step for such studies. The FQH edge structures in a confined region are typically presupposed to be the same as that in the open region in analysis of experimental results, but whether they remain unchanged with extra confinement is obscure. In this work, we present a series of unexpected plateaus in a confined single-layer two-dimensional electron gas (2DEG), which are quantized at anomalous fractions such as 9/4, 17/11, 16/13 and the reported 3/2. We explain all the plateaus by assuming surprisingly larger filling factors in the confined region. Our findings enrich the understanding of edge states in the confined region and in the applications of gate manipulation, which is crucial for the experiments with quantum point contact and interferometer.

In two-dimensional systems, in particular fractional quantum Hall (FQH) states, quasi-particles named anyons with fractional statistics can obey neither fermion nor boson statistics. Fascinatingly, quasi-particles in some FQH states may also obey non-Abelian statistics, which facilitate fault-tolerant topological quantum computation[1,2]. The FQH effect emerges from the electron–electron interaction, and cannot be treated with perturbation theory. When the bulk states are insulated, gapless excitations flow in one-dimensional channels along the boundary, and are named edge modes. The fundamental difference of FQH states is not only represented in different quantization values of the Hall resistance, the famous von Klitzing constant $R_K(h/e^2)$ divided by fractions, but also hidden in the difference of edge structures. Accumulating knowledge of the edge has been learned from various experimental methods, and lateral confinement, which brings edges in proximity, has been widely used[3–13]. Weak-tunneling experiments have been exploited to derive effective charge and interaction strength, and thus to distinguish between candidates of the 5/2 FQH state[14–16]. Interference attempts have been progressed by multiple positions of edge weak-tunneling[6,7,13]. Noise measurement techniques have been developed to probe effective charge[4,5], neutral modes[8], and quantum statistics[12] while thermal transport has been used to distinguish between Abelian and non-Abelian topological orders[17].

In these lateral confinement experiments, the same set of edges is expected to extend from source to drain. Thus, weak tunneling between edge states is considered at quantum point contacts. Multiple theories have been proposed to describe FQH edge modes, since conditions vary with geometry constraint, potential profile, disorder, and screening. For hole-conjugate states, e.g., the $\nu = 2/3$ state, the edge structure was predicted to be one integer mode accompanied by one counter-propagating −1/3 fractional mode[18]. Considering the role of disorder, edge modes were also proposed to be one 2/3 downstream mode and one upstream neutral mode[19,20]. The transition of edge structures indicated by the two-terminal conductance has been observed in synthetic edge modes[21]. Nevertheless, observations of an extra $(1/3)\cdot(e^2/h)$ plateau with quantum point contacts[22,23] pointed out the complexity of edge structure, which might be consistent with edge modes consisting of four brunches $-1/3, +1, -1/3, +1/3$ under shallow confining potential[24,25]. The FQH states with the simplest denominator three provide an example of how the edge structure can be affected in a realistic device. Recently, our group and a group from Tohoku University observed an anomalous even denominator $(3/2)\cdot(e^2/h)$ plateau in the 5/3 FQH state in the devices with confined structure[26,27], and the edge structure accomplished with the 3/2 plateau is of curiosity. It is hard to believe that the edge modes play no role in the formation of such an unexpected quantization.

[1]International Center for Quantum Materials, Peking University, Beijing 100871, China. [2]Department of Electrical Engineering, Princeton University, Princeton, NJ 08544, USA. [3]School of Physics, Zhejiang University, Hangzhou 310027, China. [4]CAS Center for Excellence in Topological Quantum Computation, University of Chinese Academy of Sciences, Beijing 100190, China. [5]Interdisciplinary Institute of Light-Element Quantum Materials and Research Center for Light-Element Advanced Materials, Peking University, Beijing 100871, China. e-mail: xilin@pku.edu.cn

To discuss the edge modes inside a confined region, uniform density is usually required as a precondition. However, the decrease of electron density in the confined region is a common phenomenon due to the fringe field of negative gate voltages. The annealing process[14–16] has been applied to modulation-doped GaAs quantum wells to obtain a sharp confining potential and, therefore, a uniform density inside and outside confined regions. The annealing, which applies a negative gate voltage at relatively high temperatures before cooling down to the actual experimental temperature[14–16], was also used in our experiments. We observed the same 3/2 plateau as we and others reported[26,27], with an additional series of anomalous fractions, such as 9/4, 17/11, and 16/13. We suggest the mechanism of these plateaus result from a surprising density enhancement in a micron-scale confined region and we propose a model to describe the propagation and reflection of edge modes inside and outside the confined region. If this model should be applied, it provides insight into the details of the edge, including the selected edge transmission and equilibration of chemical potentials. Accompanied by the reliable determination of quantization fractions, the confinement in this experiment provides a wedge to probe the FQH edge behaviors.

## Results

Devices for this experiment are fabricated into a Hall bar geometry with gate-defined lateral confinement (Fig. 1a). The pair of contacts adjacent to the split gates measure the diagonal resistance $R_D$ across the confined region, while the other pair measure the Hall resistance $R_{XY}$ far away from the constriction in the open region. Electrons underneath the gates can be depleted at $V_g \approx -1.3$ V, leaving only a narrow conductive channel between the split gates, which is indicated as a step-like increase of $R_D$ in Fig. 1b. More negative voltages shrink the channel further, exhibited as a slow increase in $R_D$. We apply $V_g$ at temperature $T = 6$ K and then cool the devices down to the lowest temperature $T<10$ mK in our dilution refrigerator with the same voltage.

Figure 1c shows $R_{XY}$ and $R_D$ traces in the filling factor range of $1<\nu<2$ (longitudinal resistance $R_{XX}$ in the open region and $R_L$ across the confined region shown in Supplementary Fig. 2). The label $\Delta V = V_g - V_g(6\,\text{K})$ denotes the variation of the gate voltage from the value at 6 K to the base temperature. The quantization values in $R_{XY}$ trace are the filling factors of the 2DEG in the open region, which depend on the 2DEG density and the magnetic field. $R_D(\Delta V = 0)$ and $R_{XY}$ (red and black traces) demonstrate similar behaviors in several integer quantum Hall (IQH) and FQH states, suggesting that the electron density in the confined region is similar to that in the open region and edge modes fully transmit through the confined region as long as we don't change the gate voltage.

However, as $V_g$ changes to less negative values subsequently, for example, from $-4.5$ V to $-1.4$ V, $R_D$ (blue trace) shows prominent lifting between $\nu = 1$ and $\nu = 2$ plateaus. When $R_{XY}$ is quantized at 5/3 FQH state, $R_D$ exhibits a plateau quantized at $R_K/(3/2)$, as reported in previous works[26,27]. And more $R_D$ plateaus come along with the 3/2 plateau at different magnetic fields, detailed traces of which are shown in Fig. 2.

Anomalous plateaus are observed at $\nu = 8/5$, 7/5, and 4/3. The well-quantized Hall resistance $R_{XY}$ is a prerequisite for the appearance of anomalous $R_D$ plateaus. The magnetic field ranges of $R_D$ and $R_{XY}$ plateaus are mostly coincident. It is worthwhile to mention that $R_D$ can switch between different quantized values with different gating conditions when $R_{XY}$ doesn't change. In Fig. 2b, $R_{XY}$ (black trace) shows the $\nu = 8/5$ FQH plateau, while $R_D$ switches from $R_K/(17/11)$ at $\Delta V = 0.5$ V (blue trace) to $R_K/(10/7)$ at $\Delta V = 2.5$ V (red trace). Experimentally, the

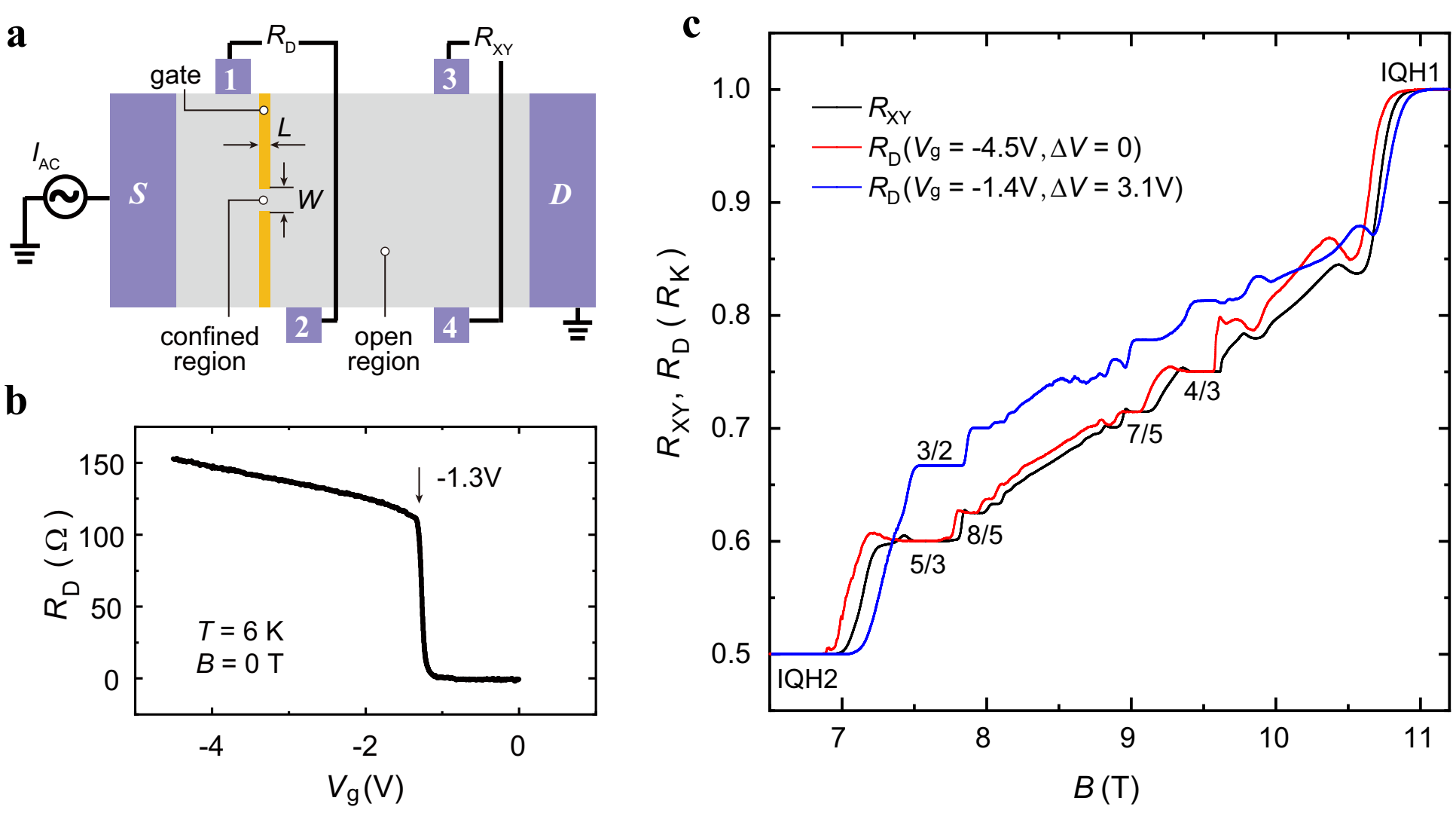

**Fig. 1 | Sample geometry, diagonal resistance $R_D$ versus gate voltage $V_g$, and $R_D$ and Hall resistance $R_{XY}$ versus magnetic field. a** The Hall bar devices are 150 μm wide and metal split gates define regions of $L = 1$ μm and $W = 2$ μm. $R_D$ measures diagonal resistance across the confined region while $R_{XY}$ measures the Hall resistance of the open region. **b** $R_D$ as a function of $V_g$ at $T = 6$K and zero magnetic field. Electrons are depleted at $V_g \approx -1.3$V, which is shown as a step-like increase in the plot. **c** $R_{XY}$ and $R_D$ as a function of the magnetic field at refrigerator temperature $T = 13$mK (in the unit of $R_K = h/e^2$). Several FQH and IQH states are developed (labeled in the figure) in $1<\nu<2$. The red trace ($\Delta V = 0$) follows the black trace in IQH and FQH states, while the blue trace ($\Delta V = 3.1$V) shows prominent rising and some unexpected plateaus. Source data are provided as a Source Data file.

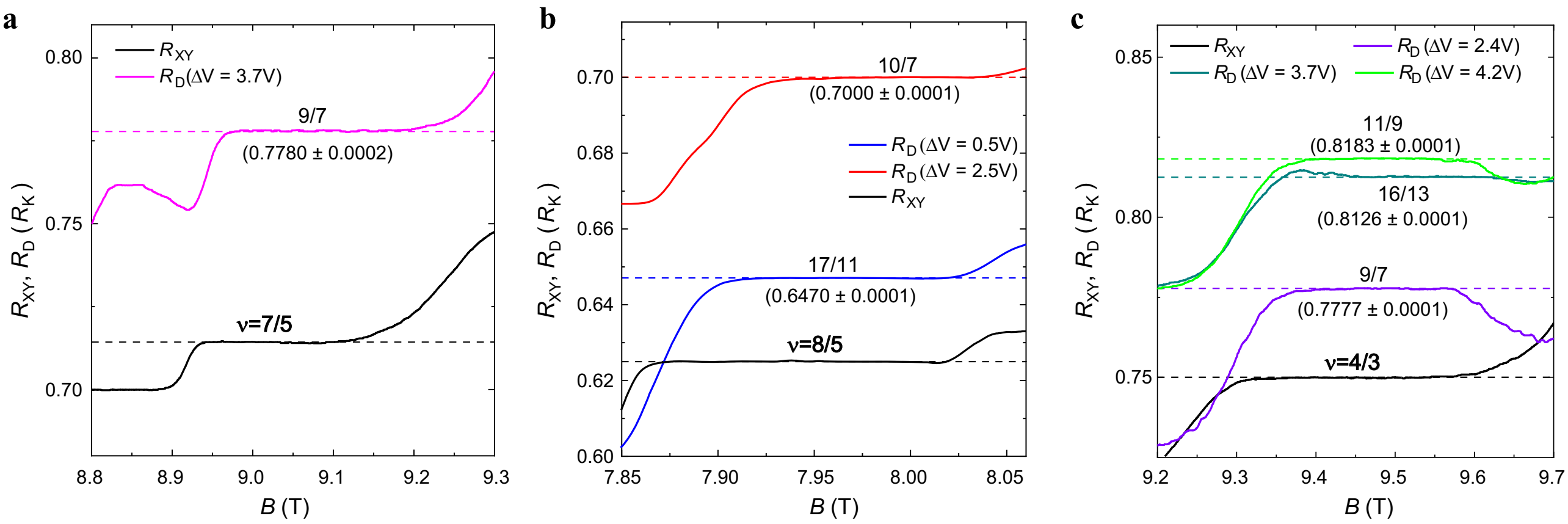

**Fig. 2 | Anomalous plateaus at $\nu = 7/5$, $8/5$, and $4/3$ states with different gate voltages.** Black traces represent $R_{XY}$ and traces in other colors represent $R_D$ at different gating conditions. The annealing voltages are −5.0 V in **a**, −4.5 V in **b**, and −5.5, −5.0, −5.0 V from top to bottom in **c**. In **b** and **c**, $R_D$ exhibits different plateaus at different $\Delta V$. The mean values and standard deviations of $R_D$ (in a range of at least 0.07 T and 110 points, depending on the particular plateau) are labeled next to the plateaus (in the unit of $R_K$), and dashed lines are guides to the eye. Source data are provided as a Source Data file.

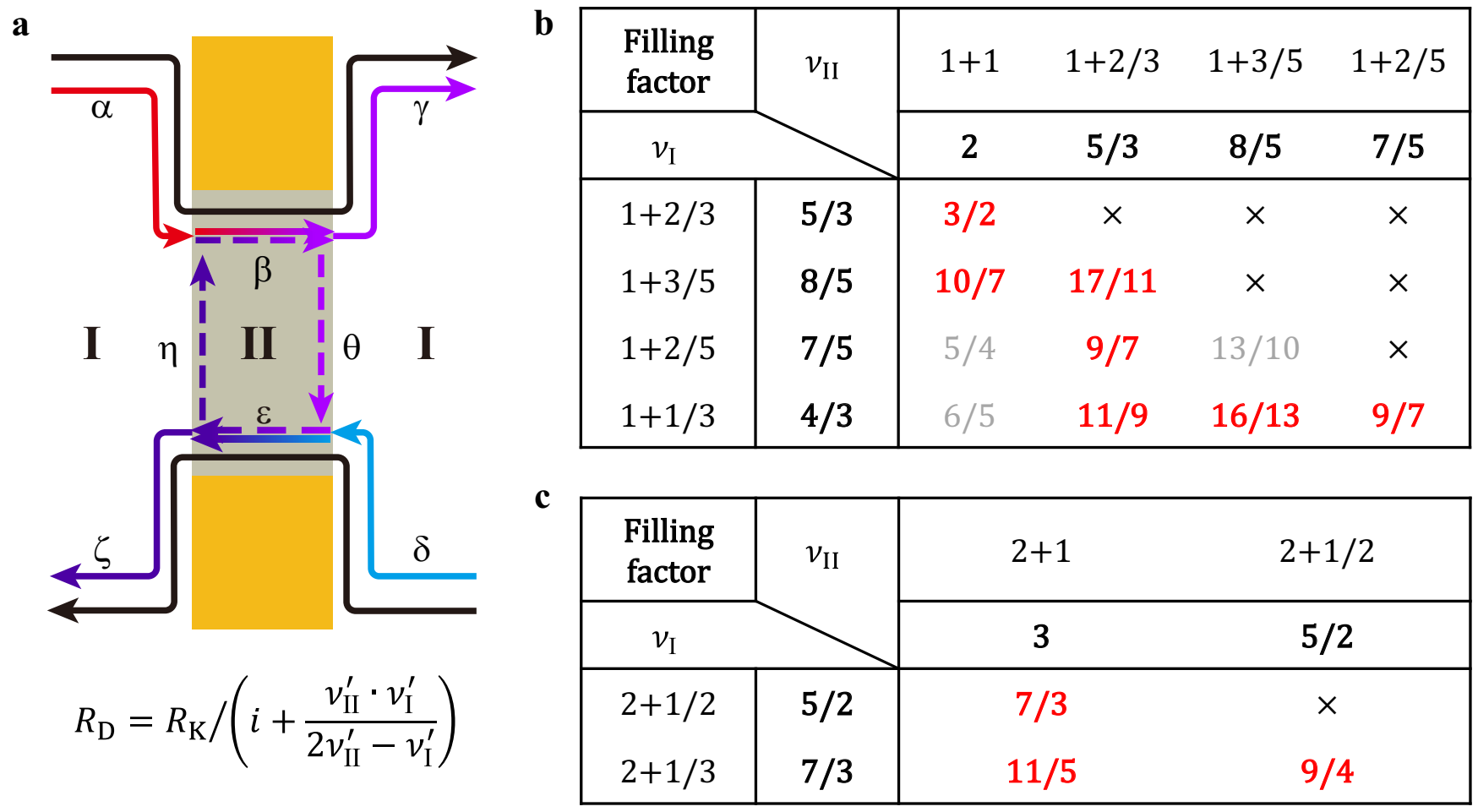

$$R_D = R_K \Big/ \left( i + \frac{\nu'_{II} \cdot \nu'_I}{2\nu'_{II} - \nu'_I} \right)$$

b

| Filling factor $\nu_I$ | $\nu_{II}$ | 1+1 | 1+2/3 | 1+3/5 | 1+2/5 |
|---|---|---|---|---|---|
| | | 2 | 5/3 | 8/5 | 7/5 |
| 1+2/3 | 5/3 | 3/2 | × | × | × |
| 1+3/5 | 8/5 | 10/7 | 17/11 | × | × |
| 1+2/5 | 7/5 | 5/4 | 9/7 | 13/10 | × |
| 1+1/3 | 4/3 | 6/5 | 11/9 | 16/13 | 9/7 |

c

| Filling factor $\nu_I$ | $\nu_{II}$ | 2+1 | 2+1/2 |
|---|---|---|---|
| | | 3 | 5/2 |
| 2+1/2 | 5/2 | 7/3 | × |
| 2+1/3 | 7/3 | 11/5 | 9/4 |

**Fig. 3 | Sketch of edge modes propagation inside and outside the confined region and predicted plateaus in $1 < \nu \le 2$ and $2 < \nu \le 3$. a** Black arrows represent outer edge modes that each contributes to $e^2/h$, while arrows in other colors represent inner edge modes contribute to a fraction (<1) multiples $e^2/h$. Red color (α) represents the highest chemical potential of the inner edge modes, while blue color (δ) represents the lowest chemical potential. Region I represents the open region and region II represents the confined region, as described in Fig. 1a. The equation is Eq. 1 in the text. **b** Predicted plateaus in $1 < \nu \le 2$ according to Eq. 1. Fractions in red bold font indicate plateaus observed in experiments while those in gray font haven't been observed. Crosses indicate cases that do not apply to the model. c Predicted plateaus in $2 < \nu \le 3$ when taking $i = 2$ in Eq. 1.

measured resistance is given as a decimal rather than a fraction. The plateaus in our experiments can be identified with high accuracy (within 0.0001 or 0.0002 in the unit of $R_K$), and the fractions determined are the most adjacent ones whose denominators are within 100 (Supplementary Fig. 4).

To understand these anomalous plateaus, we consider the edge structures inside and outside the confined region, and their behaviors at the interface together. The experimental device is schematically simplified by the model in Fig. 3a. Region I represents the open region of the device, while region II represents the gate-defined confined region. When parts of the edge modes are reflected directly from the region I to region II, $R_D$ will show plateaus differing from $R_{XY}$. However, such a trivial picture cannot explain the appearance of anomalous plateaus such as the $R_K/(3/2)$ plateau emergent from the $\nu = 5/3$ FQH state (Supplementary Note 1 and Supplementary Fig. 1). Here, we propose a model to explain all the plateaus systematically. The model is based on two assumptions.

First, the electron density in region II is larger than that in region I (i.e., $n_{II} > n_I$). Before our measurement, we start from a uniform density for both regions I and II (red trace in Fig. 1c), and then substantially charge the gates to less negative voltages at lower temperatures, i.e., $\Delta V > 0$. Therefore, it is reasonable to assume a density enhancement in the later procedure, which leads to $n_{II} > n_I$. This is also consistent with the results in ref. [27] where the 3/2 plateau emerges when an extra gate with a positive voltage covers the confined region[27]. The density mismatch leads to different filling factors between the two regions. When electrons in regions I and II are in different incompressible states, our devices form $\nu_I - \nu_{II} - \nu_I$ junctions with $\nu_{II} > \nu_I$, although the common side-effect of the lateral gate-defined confinement is the density reduction in the area around gates. And edge modes in the two regions are different from each other.

Second, behaviors of integer and fractional edge modes differ when they propagate through the $\nu_I - \nu_{II} - \nu_I$ junction. As the variation between $n_I$ and $n_{II}$ is not significant, the integer edge modes in the region I transmit through the confinement directly. For example, in the 4/3 FQH state, the edge modes are taken as $\nu_I = i + \nu_I' = 1 + 1/3$, separated into two parts: one IQH edge mode contributes to $e^2/h$ in the outermost, and the other edge mode contributes to $(1/3) \cdot (e^2/h)$. The outermost IQH edge mode (black arrows in Fig. 3a) exhibits full transmission through the $\nu_I - \nu_{II} - \nu_I$ junction, as long as region II is in an incompressible state in $1 < \nu \le 2$. It propagates through the confined region without dissipation. In other words, inside the confined region, there is no inter-channel tunneling between outer edge modes and inner edge modes, or scattering between outer edge modes and the bulk states. Thus, each IQH edge in region I contributes to $e^2/h$. The fractional edge modes have more complicated behaviors. There is no reflection from region I to region II (α or δ), as $\nu_{II}$ is larger than $\nu_I$. However, edge modes traversing from region II to region I (β or ε) are partially reflected as a result of the filling factor difference[28]. The reflected edge currents (η or θ) propagate along the interfaces of two regions and are mixed with edge modes transmitted from the region I (α or δ). These edge modes carry different chemical potentials and obtain equilibration inside region II.

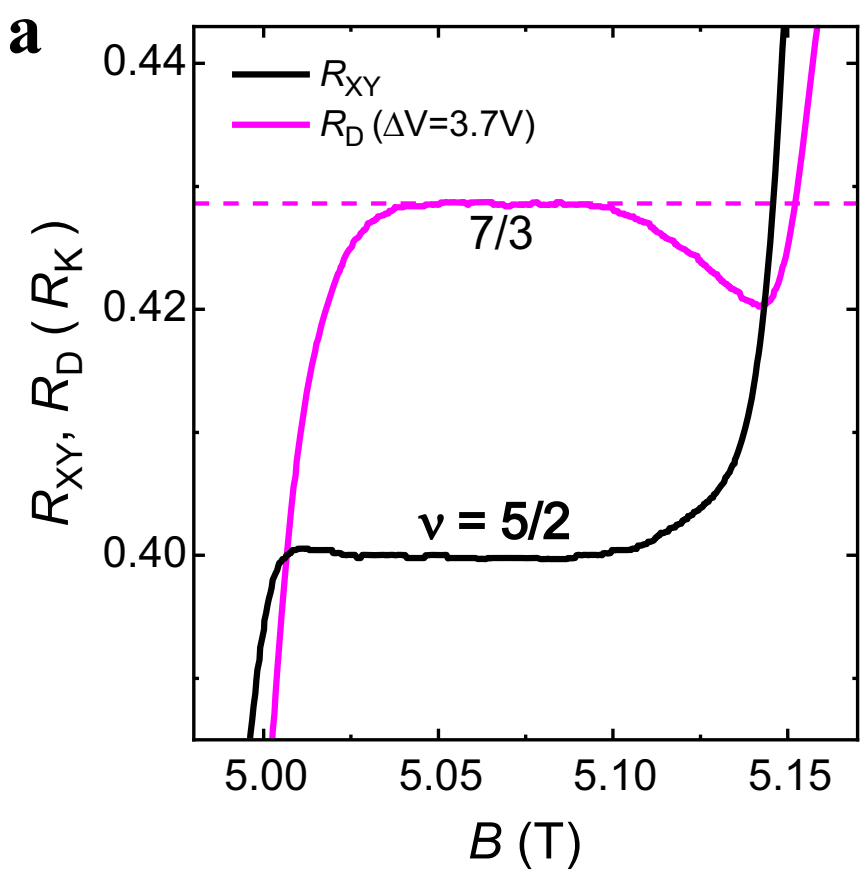

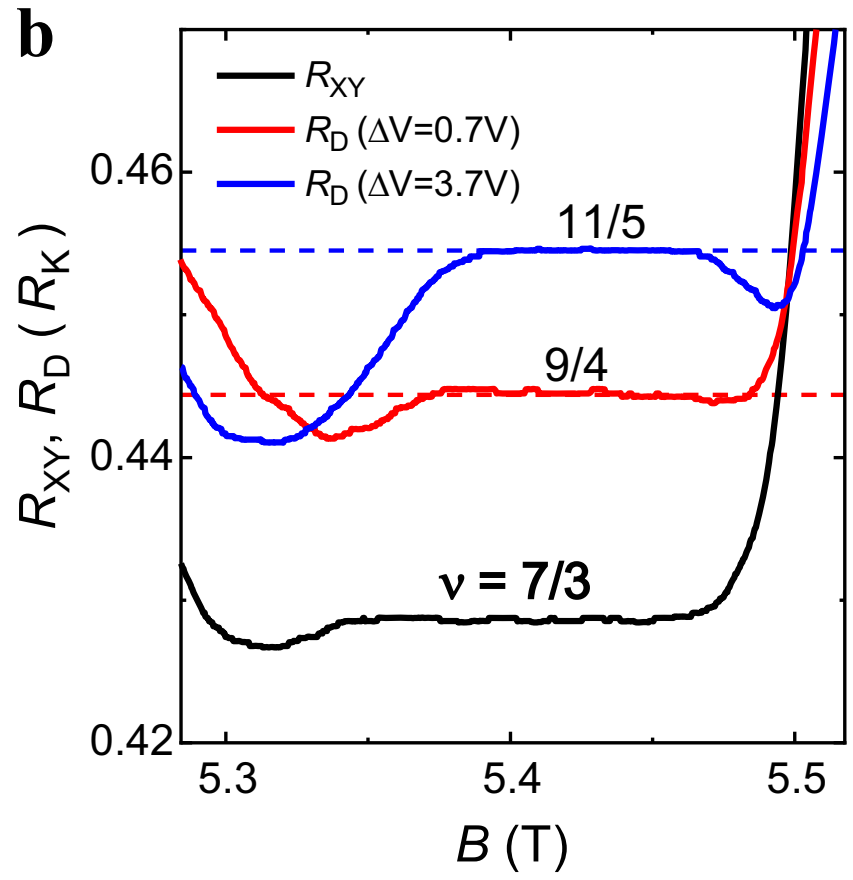

**Fig. 4 | Anomalous plateaus at $\nu = 5/2$ and $\nu = 7/3$ with different gate voltages.** Black traces represent $R_{XY}$ and traces in other colors represent $R_D$ at different gating conditions. All the plateaus are consistent with predicted values in Fig. 3c. The annealing voltages are −5.0 V in **a**, −5.0 V (blue) and −2.0 V (red) in **b**. Source data are provided as a Source Data file.

Based on this model, the diagonal resistance of the $\nu_I - \nu_{II} - \nu_I$ junction ($\nu_{II} > \nu_I$) is derived as (Supplementary Note 1)

$$R_D = R_K / \left( i + \frac{\nu_{II}' \cdot \nu_I'}{2\nu_{II}' - \nu_I'} \right) \tag{1}$$

where $i = 1$ represents the integer edge mode, $\nu_{I,II}' = \nu_{I,II} - i$ represents the rest of the edge modes, and $0 < \nu_{I,II}' \le 1$. Based on Eq. 1, we summarize the expected quantized fractions for $\nu_I = 5/3, 8/5, 7/5, 4/3$ and $\nu_{II} = 2, 5/3, 8/5, 7/5$ in Fig. 3b. All the red fractions have been observed in Figs. 1c, 2. For a given $\nu_I$, $R_D$ increases with larger $\Delta V$ experimentally (Fig. 2b, c), which is consistent with our assumption that the electron density in region II increases with increasing $\Delta V$ (Supplementary Note 3 and Supplementary Fig. 3). The $\nu_I - \nu_{II} - \nu_I$ junction equilibration model agrees well with our experimental results.

Edge equilibration experiments have been investigated in some quantum Hall states[29–35]. Here we construct 1-µm-size $\nu_I - \nu_{II} - \nu_I$ junctions ($\nu_{II} > \nu_I$) using split-gate-defined lateral confinement and demonstrate selective equilibration of fractional edge modes. We consider behaviors of many-body fractional edges together with single-particle integer edges at the same time. And well-quantized resistance allows us to understand behaviors of edge modes with lateral confinement.

To further prove our model, we apply it to FQH states with multiple integer edges. Predicted plateaus in $2 < \nu_I < 3$ with $i = 2$ in Eq. (1) are listed in Fig. 3c, which agree with experimental observations (Fig. 4a, b). At $\nu_I = 5/2$, the observed plateau $R_K/(7/3)$ indicates that a 1/2 edge mode ($\nu_I' = 1/2$) is spatially separated from two integer edge modes and gets equilibrium with $\nu_{II}' = 1$ in the confined region. At $\nu_I = 7/3$, predicted quantization values of $R_K/(11/5)$ and $R_K/(9/4)$ are both observed. The fraction 9/4 has no connection to a new even denominator FQH state, but only indicates the existence of a stable 5/2 FQH state in region II at a restricted dimension of ~1 µm.

## Discussion

Charge edge modes in FQH states, including in hole-conjugate states, are treated as a combination of co-propagating integer and fractional edge modes in our model. In the scenario described in Fig. 3a, the $\nu = 5/3$ edge mode consists of two parts: the outer integer part and the inner fractional part. And both parts flow exclusively downstream. Similarly, the charge modes in $\nu = 5/2$ FQH state in typical quantum point contacts are likely to be two downstream integer modes plus fractional downstream 1/2 mode, as expected. All the hole-conjugate states that appear in both region I and region II are described as $i$ integer modes and a set of downstream fractional modes for current carrying in our model. It is worth mentioning that the combination of $(i + 1)$ integer modes and upstream fractional modes is also possible for hole-conjugate states. The investigation of such interaction requires an in-depth theoretical study of the additional inter-edge tunneling, which is beyond this experimental report.

The mismatch density-induced quantization in this work is an approach to probe edge behaviors. The density mismatch junction serves as an artificial contact, which only equilibrates fractional modes and leaves integer modes to pass through freely at temperatures of about 12 mK[36]. If this selective chemical potential equilibration can be extended to different channels among FQH edge modes, fine structures of edge modes might be determined. For example, the edge structures of $\nu = 2/3$ and $\nu = 2/5$ FQH states have been proposed to be $2/3 = 1/3 + 1/3$ and $2/5 = 1/3 + 1/15$, respectively[23,37,38]. If the outer 1/3 edge mode traverses the confined region freely as integer edges in our model and leaves only inner modes to equilibrate, edge structures will be revealed by the new quantized plateau.

Modeling the plateaus in our measurements requires two experimental conditions in this work. Firstly, a moderate sensitivity in resistance measurement is necessary in order to correctly determine the fraction from a measured decimal. We have demonstrated that a conventional lock-in technique is satisfactory to identify fractions with denominators less than 100. Secondly, a higher density in a confined region is involved. Such a higher density region is created by negative voltage counter-intuitively, through gate voltage variation sequence as a function of temperature.

Additionally, chemical potential equilibrium exists only in the inner edge modes in the scale of our device, concluded from the quantized $R_D$ values. There is no tunneling between the outer integer modes and the inner fractional modes or scattering between the insulating bulk state and edges in the confined region. The order of 1 μm is a typical dimension in recent edge experiments with gate confinements[6–8,12,14–16,23,26,27,39], due to the depth of 2DEG, the convenience of fabrication and breakdown effect of FQH states[40]. It is comfortable to know the inner fractional modes have no equilibration with outer integer modes, which is hidden in previous experimental analyses[6,14–16]. And if the density mismatch disappears, interactions between fractional edges could be suppressed and edge modes could propagate a longer distance, which will be more suitable for interference experiments. Lastly, we would like to note that the theoretical consideration of the density mismatch situation is similar to that of Fig. 3a, with $\nu_I = 2/3$ and $\nu_{II} = 1$, suggested ${R_D}^{-1} = 1/2\, e^2/h$ under certain conditions due to the partial reflection at the interface[41].

In summary, we observe a series of anomalous quantized plateaus and explain them with an elevated density in the confined region. FQH states can survive in the dimension of around 1 μm, including the 5/2 state for searching signals of non-Abelian statistics in interferometers. The selected edge transmission is exhibited by a series of apparent resistance plateaus. Such an approach leads to a better understanding of FQH edge behaviors and interpretation of edge experiments.

## Methods

### Sample fabrication

Devices are fabricated from a GaAs/AlGaAs heterostructures wafer with single-layer two-dimensional electron gas buried 200 nm under the surface. These devices are not exposed to LED after cooling down. The densities of the devices are around $3.0 \times 10^{11}\,\mathrm{cm}^{-2}$ and the mobilities are above $10^7\,\mathrm{cm}^2\,\mathrm{V}^{-1}\,\mathrm{s}^{-1}$ at low temperatures. The devices are wet-etched into Hall bar geometry after UV lithography, and the gates are deposited of Cr/Au after electron beam lithography. Ohmic contacts are made from annealed Pt/Au/Ge.

### Measurement techniques

The resistance is measured by standard lock-in techniques at 6.47 or 17 Hz with 1 nA. The measurements are carried out in a dilution refrigerator with a base refrigerator temperature below 6 mK. Thermocoax cables and cryogenic RC filters are connected to each lead of the devices, to ensure a negligible temperature difference between electrons and the refrigerator above 12 mK. Before cooling down the devices to the lowest temperature, we apply appropriate gate voltages at 6 K and wait for more than 6 h. The data presented in this work are from two devices with the same geometry, and the key observations can be reproduced in other devices and separate cool-downs.

## Data availability

The authors declare that the main data supporting the findings of this study are available. Extra data were available from the corresponding author upon request. Source data are provided with this paper.

## References

1. Kitaev, A. Y. Fault-tolerant quantum computation by anyons. *Ann. Phys.* **303**, 2–30 (2003).
2. Nayak, C., Simon, S. H., Stern, A., Freedman, M. & Das Sarma, S. Non-Abelian anyons and topological quantum computation. *Rev. Mod. Phys*. **80**, 1083–1159 (2008).
3. Heiblum, M. & Feldman, D. E. *Fractional Quantum Hall Effects* (eds Halperin, B. I. & Jain, J. K.) Ch. 4 (World Scientific Publishing Company, 2020).
4. de-Picciotto, R. et al. Direct observation of a fractional charge. *Nature* **389**, 162–164 (1997).
5. Saminadayar, L., Glattli, D. C., Jin, Y. & Etienne, B. Observation of the e/3 fractionally charged Laughlin quasiparticle. *Phys. Rev. Lett.* **79**, 2526–2529 (1997).
6. Willett, R. L., Pfeiffer, L. N. & West, K. W. Measurement of filling factor 5/2 quasiparticle interference with observation of charge e/4 and e/2 period oscillations. *Proc. Natl Acad. Sci. USA* **106**, 8853–8858 (2009).
7. Zhang, Y. et al. Distinct signatures for Coulomb blockade and Aharonov-Bohm interference in electronic Fabry-Perot interferometers. *Phys. Rev. B* **79**, 241304 (2009).
8. Bid, A. et al. Observation of neutral modes in the fractional quantum Hall regime. *Nature* **466**, 585–590 (2010).
9. Willett, R. L. The quantum Hall effect at 5/2 filling factor. *Rep. Prog. Phys.* **76**, 076501 (2013).
10. Lin, X., Du, R. R. & Xie, X. C. Recent experimental progress of fractional quantum Hall effect: 5/2 filling state and graphene. *Natl Sci. Rev.* **1**, 564–579 (2014).
11. Jiang, N. & Wan, X. Recent progress on the non-Abelian $\nu$ = 5/2 quantum Hall state. *AAPPS Bull.* **29**, 58 (2019).
12. Bartolomei, H. et al. Fractional statistics in anyon collisions. *Science* **368**, 173–177 (2020).
13. Nakamura, J., Liang, S., Gardner, G. C. & Manfra, M. J. Direct observation of anyonic braiding statistics. *Nat. Phys.* **16**, 931–936 (2020).
14. Radu, I. P. et al. Quasi-particle properties from tunneling in the $\nu$ = 5/2 fractional quantum Hall state. *Science* **320**, 899–902 (2008).
15. Lin, X., Dillard, C., Kastner, M. A., Pfeiffer, L. N. & West, K. W. Measurements of quasiparticle tunneling in the $\nu$ = 5/2 fractional quantum Hall state. *Phys. Rev. B* **85**, 165321 (2012).
16. Fu, H. et al. Competing $\nu$ = 5/2 fractional quantum Hall states in confined geometry. *Proc. Natl Acad. Sci. USA* **113**, 12386–12390 (2016).
17. Banerjee, M. et al. Observation of half-integer thermal Hall conductance. *Nature* **559**, 205–210 (2018).
18. MacDonald, A. H. Edge states in the fractional-quantum-Hall-effect regime. *Phys. Rev. Lett.* **64**, 220–223 (1990).
19. Kane, C. L., Fisher, M. P. & Polchinski, J. Randomness at the edge: theory of quantum Hall transport at filling $\nu$ = 2/3. *Phys. Rev. Lett.* **72**, 4129–4132 (1994).
20. Kane, C. L. & Fisher, M. P. Impurity scattering and transport of fractional quantum Hall edge states. *Phys. Rev. B* **51**, 13449–13466 (1995).
21. Cohen, Y. et al. Synthesizing a $\nu$ = 2/3 fractional quantum Hall effect edge state from counter-propagating $\nu$ = 1 and $\nu$ = 1/3 states. *Nat. Commun.* **10**, 1920 (2019).
22. Bid, A., Ofek, N., Heiblum, M., Umansky, V. & Mahalu, D. Shot noise and charge at the 2/3 composite fractional quantum Hall state. *Phys. Rev. Lett.* **103**, 236802 (2009).
23. Sabo, R. et al. Edge reconstruction in fractional quantum Hall states. *Nat. Phys.* **13**, 491–496 (2017).
24. Meir, Y. Composite edge states in the $\nu$ = 2/3 fractional quantum Hall regime. *Phys. Rev. Lett.* **72**, 2624–2627 (1994).
25. Wang, J., Meir, Y. & Gefen, Y. Edge reconstruction in the $\nu$ = 2/3 fractional quantum Hall state. *Phys. Rev. Lett.* **111**, 246803 (2013).
26. Fu, H. et al. 3/2 fractional quantum Hall plateau in confined two-dimensional electron gas. *Nat. Commun.* **10**, 4351 (2019).
27. Hayafuchi, Y. et al. Even-denominator fractional quantum Hall state in conventional triple-gated quantum point contact. *Appl. Phys. Express* **15**, 025002 (2022).
28. Abanin, D. A. & Levitov, L. S. Quantized transport in graphene p-n junctions in a magnetic field. *Science* **317**, 641–643 (2007).
29. Muller, G. et al. Equilibration length of electrons in spin-polarized edge channels. *Phys. Rev. B* **45**, 3932–3935 (1992).

30. Ozyilmaz, B. et al. Electronic transport and quantum hall effect in bipolar graphene p-n-p junctions. *Phys. Rev. Lett.* **99**, 166804 (2007).
31. Amet, F., Williams, J. R., Watanabe, K., Taniguchi, T. & Goldhaber-Gordon, D. Selective equilibration of spin-polarized quantum Hall edge states in graphene. *Phys. Rev. Lett.* **112**, 196601 (2014).
32. Yang, J. W. et al. Equilibration and filtering of quantum Hall edge states in few-layer black phosphorus. *Phys. Rev. Mater.* **4**, 114008 (2020).
33. Nicolí, G. et al. Spin-selective equilibration among integer quantum Hall edge channels. *Phys. Rev. Lett.* **128**, 056802 (2022).
34. Maiti, T. et al. Magnetic-field-dependent equilibration of fractional quantum Hall edge modes. *Phys. Rev. Lett.* **125**, 076802 (2020).
35. Maiti, T. et al. Temperature-dependent equilibration of spin orthogonal quantum Hall edge modes. *Phys. Rev. B* **104**, 085304 (2021).
36. Sun, J. et al. Dynamic ordering transitions in charged solid. *Fundam. Res.* **2**, 178–183 (2022).
37. Beenakker, C. W. J. Edge channels for the fractional quantum Hall effect. *Phys. Rev. Lett.* **64**, 216–219 (1990).
38. Chang, A. M. & Cunningham, J. E. Transport evidence for phase separation into spatial regions of different fractional quantum Hall fluids near the boundary of a two-dimensional electron gas. *Phys. Rev. Lett.* **69**, 2114–2117 (1992).
39. Banerjee, M. et al. Observed quantization of anyonic heat flow. *Nature* **545**, 75–79 (2017).
40. Dillard, C., Lin, X., Kastner, M. A., Pfeiffer, L. N. & West, K. W. Breakdown of the integer and fractional quantum Hall states in a quantum point contact. *Phys. E Low. Dimens. Syst. Nanostruct.* **47**, 290–296 (2013).
41. Lai, H.-H. & Yang, K. Distinguishing particle-hole conjugated fractional quantum Hall states using quantum-dot-mediated edge transport. *Phys. Rev. B* **87**, 125130 (2013).

## Acknowledgements

We thank Bo Yang, Xin Wan, Junren Shi, Rui-rui Du, and Yang Xu for their discussions. The work at PKU was supported by the National Key Research and Development Program of China (2021YFA1401901), the NSFC (12141001 and 11921005), and the Strategic Priority Research Program of the Chinese Academy of Sciences (Grant No. XDB28000000). The work at Princeton University was funded by the Gordon and Betty Moore Foundation through the EPiQS initiative Grant GBMF4420, by the National Science Foundation MRSEC Grant DMR-1420541, and by the Keck Foundation.

## Author contributions

J.Y., S.Y., and H.F. performed the measurements. H.F. fabricated the devices; L.N.P. and K.W.W. provided the GaAs 2DEG wafer. J.Y., X.Liu, H.F., and X.Lin analysed the data. J.Y., Y.W., X.C.X., and X.Lin developed the model and discussed it with Y.L. and H.F. J.Y. and X.Lin wrote the manuscript with input from all authors. X.Lin supervised the project.

## Competing interests

The authors declare no competing interests.

## Additional information

**Supplementary information** The online version contains supplementary material available at https://doi.org/10.1038/s41467-023-37495-9.

**Correspondence** and requests for materials should be addressed to Xi Lin.

**Peer review information** *Nature Communications* thanks Yutao Li, and the other, anonymous, reviewers for their contribution to the peer review of this work.